\documentclass[aps,prd,twocolumn,showpacs,groupedaddress,preprintnumbers,superscriptaddress,amsmath,amssymb,floatfix,longbibliography]{revtex4-1}

\usepackage[colorlinks=true, filecolor=blue, citecolor=blue,urlcolor=blue]{hyperref}
\usepackage{url}

\usepackage{bm,pifont}
\usepackage{amssymb}
\usepackage{amsmath}
\usepackage{tabularx}
\usepackage{amsfonts}
\usepackage{multirow}
\usepackage{tabularx}
\usepackage{booktabs}
\usepackage{epsfig}
\usepackage{graphicx}
\usepackage{multirow}
\usepackage{tabularx}
\usepackage{color}
\usepackage{slashed}
\usepackage{orcidlink}
\newcommand{\seq}{\begin{subequations}}
\newcommand{\sen}{\end{subequations}}
\newcommand{\eq}{\begin{eqnarray}}
\newcommand{\en}{\end{eqnarray}}

\def\shiftdown#1{#1\llap{\lower.04ex\hbox{#1}}}

\begin{document}

\title{Missing energy signatures of inelastic magnetic dipole DM at NA64e} 

%\title{Constraining dark axion portal from missing momentum signatures and EDM data of fermions} 

\author{Sergei N.~Gninenko \orcidlink{0000-0001-6495-7619}} 
%\email[\textbf{e-mail}: ]{Sergei.Gninenko@cern.ch}
\affiliation{Institute for Nuclear Research, 117312 Moscow, Russia}
\affiliation{Bogoliubov Laboratory of Theoretical Physics, JINR, 141980 Dubna, Russia} 
\affiliation{Millennium Institute for Subatomic Physics at
the High-Energy Frontier (SAPHIR) of ANID, \\
Fern\'andez Concha 700, Santiago, Chile}

\author{ N.~V.~Krasnikov \orcidlink{0000-0002-8717-6492}}
%\email[\textbf{e-mail}: ]{nikolaykrasnikov6@gmail.com}
\affiliation{Institute for Nuclear Research, 117312 Moscow, Russia}
\affiliation{Bogoliubov Laboratory of Theoretical Physics, JINR, 141980 Dubna, Russia} 

%\author{Valery~E.~Lyubovitskij \orcidlink{0000-0001-7467-572X}}
%\email[\textbf{e-mail}: ]{valeri.lyubovitskij@uni-tuebingen.de}
%\affiliation{Institut f\"ur Theoretische Physik, Universit\"at T\"ubingen, \\
%Kepler Center for Astro and Particle Physics, \\ 
%Auf der Morgenstelle 14, D-72076 T\"ubingen, Germany} 
%\affiliation{Millennium Institute for Subatomic Physics at
%the High-Energy Frontier (SAPHIR) of ANID, \\
%Fern\'andez Concha 700, Santiago, Chile}

%\author{Sergey~Kuleshov~\orcidlink{0000-0002-3065-326X}}
%\email[\textbf{e-mail}: ]{sergey.kuleshov@unab.cl}
%\affiliation{Millennium Institute for Subatomic Physics at
%the High-Energy Frontier (SAPHIR) of ANID, \\
%Fern\'andez Concha 700, Santiago, Chile}
%\affiliation{Center for Theoretical and Experimental Particle Physics,
%Facultad de Ciencias Exactas, Universidad Andres Bello,
%Fernandez Concha 700, Santiago, Chile}

%\author{Alexey~S.~Zhevlakov \orcidlink{0000-0002-7775-5917}} 
%\email[\textbf{e-mail}: ]{zhevlakov@theor.jinr.ru}
%\affiliation{Bogoliubov Laboratory of Theoretical Physics, JINR, 141980 Dubna, Russia} 
%\affiliation{Matrosov Institute for System Dynamics and 
%	Control Theory SB RAS, \\  Lermontov str., 134, 664033, Irkutsk, Russia } 

\author{I.~V.~Voronchikhin \orcidlink{0000-0003-3037-636X}}
%\email[\textbf{e-mail}: ]{i.v.voronchikhin@gmail.com}
\affiliation{Institute for Nuclear Research, 117312 Moscow, Russia}
\affiliation{ Tomsk Polytechnic University, 634050 Tomsk, Russia}

\author{D.~V.~Kirpichnikov \orcidlink{0000-0002-7177-077X}}
\email[\textbf{e-mail}: ]{dmbrick@gmail.com}
\affiliation{Institute for Nuclear Research, 117312 Moscow, Russia}

\begin{abstract}
Some extensions of the Standard Model consider inelastic dark matter (iDM) as an attractive candidate for sub-GeV DM of thermal origin
 that could be detected at modern accelerators.
In the present paper, we calculate the production rate of iDM pairs  $\chi_{1} \bar{\chi}_0$ interacting with the ordinary photon via dipole magnetic moment  in the 
reaction of high-energy electron scattering on nuclei,  $e^- N \to e^- N  \chi_{1} \bar{\chi}_0$,  in the  NA64e experiment at the CERN SPS. 
We derive the projected sensitivity 
of NA64e to such particles assuming  of  $\simeq10^{13}$ 100 GeV electrons 
on  target.  We also show, that  incorporating heavy vector meson 
decays,  $\gamma^* N \to N V (\to \chi_1 \bar{\chi}_0)$, alongside bremsstrahlung-like emission of 
inelastic  dark matter pairs, $e^- N \to e^- N \gamma^* (\to \chi_1 \bar{\chi}_0)$, will allow  NA64e  to probe previously unexplored regions of the iDM parameter space, in particular
for modest mass splittings, $\Delta \simeq 5 \times 10^{-2}$, and relatively light  masses, $m_{\chi_0} \lesssim 100~\mbox{MeV}$.
\end{abstract}

\maketitle

\section{Introduction}

Over the past few decades, compelling evidence from astrophysics has confirmed 
the presence of dark matter (DM) in the Universe~\cite{Bertone:2016nfn,Bergstrom:2012fi}. This evidence is drawn from multiple 
observational phenomena, such as the rotation curves of galaxies, fluctuations 
in the cosmic microwave background, and the effects of gravitational lensing 
\cite{Cirelli:2024ssz,Bertone:2004pz,Gelmini:2015zpa}. According to contemporary cosmological data~\cite{Planck:2015fie,Planck:2018vyg}, DM makes up roughly 85\% of 
all matter in the Universe, a constituent that remains unaccounted for by the 
Standard Model (SM). 

The inelastic DM (iDM) scenario was  introduced~\cite{Tucker-Smith:2001myb,Tucker-Smith:2004mxa,Chang:2008gd} to account for unusual signals detected by the DAMA collaboration~\cite{Bernabei:2013xsa}. Since then, it has developed into a compelling theoretical framework for sub-GeV thermal DM that couples through a SM photon~\cite{Barducci:2024nvd,Jodlowski:2023ohn,Dienes:2023uve,Eby:2023wem}, hidden vector ~\cite{Garcia:2024uwf,Cui:2009xq,Jordan:2018gcd,Berlin:2018pwi,Batell:2021ooj,Mongillo:2023hbs,Izaguirre:2015zva,Foguel:2024lca,Filimonova:2022pkj,CarrilloGonzalez:2021lxm}, and scalar~\cite{Voronchikhin:2025eqm,Wang:2025xoq,DallaValleGarcia:2024zva} mediators.

Current and projected experimental constraints on iDM particles motivate several distinct search 
strategies across different facilities: (i) $e^+e^-$ collider probes at 
LEP~\cite{L3:2003yon,Fortin:2011hv} and BaBar~\cite{Izaguirre:2015zva}; (ii) past electron beam dump experiment E137~\cite{Berlin:2018pwi,Batell:2014mga}  (iii) proton beam dump 
experiments, including the past CHARM~\cite{CHARM:1985anb,Gninenko:2012eq,CHARM:1983ayi} and 
NuCal~\cite{Blumlein:1990ay,Blumlein:2011mv} facilities, as well as the proposed 
SHIP~\cite{Alekhin:2015byh} experiment;  (iv) far-forward detectors at the LHC, exemplified by 
FASER~\cite{Feng:2017uoz,Jodlowski:2023ohn,Dienes:2023uve}; (v) searches at  high intensity frontier  with the  proposed MATHUSLA~\cite{Chou:2016lxi,Curtin:2018mvb,Jodlowski:2023ohn}, 
FLArE~\cite{Batell:2021blf,Jodlowski:2023ohn}, SND@LHC~\cite{Boyarsky:2021moj}, CODEX-b~\cite{Gligorov:2017nwh,Berlin:2018jbm}, and  FORMOSA~\cite{Foroughi-Abari:2020qar} experimental facilities; and (vi) the Large Hadron collider probes of iDM  at ATLAS~\cite{Izaguirre:2015zva,Pierce:2017taw}, CMS~\cite{CMS:2018rea,CMS:2014jvv}, and LHCb~\cite{LHCb:2017trq,Ilten:2016tkc} are presented in Ref.~\cite{Berlin:2018jbm}.

In this paper, we study the missing energy signature of iDM at the electron fixed-target experiment NA64e~\cite{Voronchikhin:2025eqm,Mongillo:2023hbs,NA64:2021acr} that arises in the scenarios incorporating  two Dirac DM fermions,  with the lighter state denoted as $\chi_{0}$, and the heavier one as $\chi_{1}$. 
Their coupling to the SM photon field is mediated by magnetic dipole moment (MDM) operator  of dimension five. Specifically,
the interactions  are encoded in the  following effective Lagrangian~\cite{Chang:2010en,Jodlowski:2023ohn,Dienes:2023uve}
\begin{equation}
\label{LagrangianMDM}
    \mathcal{L} \supset \frac{1}{\Lambda_{\rm M}} \bar{\chi}_1 \sigma^{\mu \nu} \chi_0 F_{\mu \nu} + \mbox{\rm h.~c.},
\end{equation}
where $\Lambda_{\rm M}$ is the typical energy scale that governs the strength  of the new physics non-renormalizable coupling, 
$F_{\mu \nu}$ is the electromagnetic strength tensor, and  $\sigma^{\mu\nu}=i [\gamma^\mu,\gamma^\nu]/2$.

% \begin{figure}[!ht]
 \begin{figure*}[t!]
\centering
\includegraphics[width=0.7\textwidth]{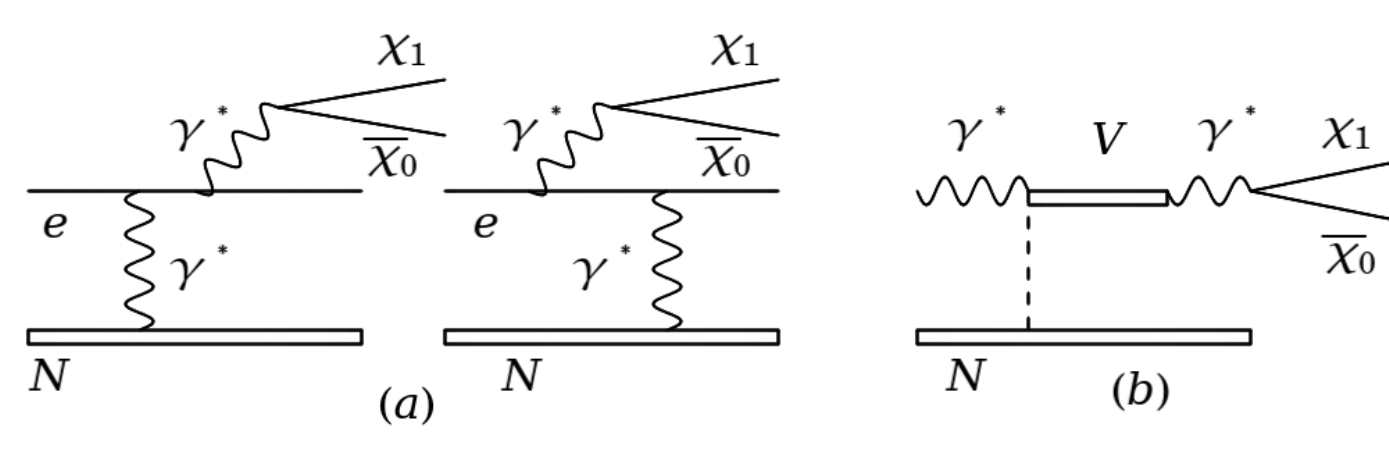}
\caption{ Feynman diagrams illustrating the production iDM pairs  in electron-nucleus
scattering for two channels: (a) bremsstrahlung like emission $e N \to e N \gamma^*(\to \chi_1 \bar{\chi}_0)$; (b) reaction of vector meson photoproduction $\gamma^* N \to N V$ followed by its radiative decay $V \to \gamma^* \to \chi_1 \bar{\chi}_0$~\cite{Barducci:2023hzo,Jodlowski:2023ohn,Dienes:2023uve}. 
\label{FeynamProduction}}
\end{figure*}

 \begin{figure}[t!]
\centering
\includegraphics[width=0.25\textwidth]{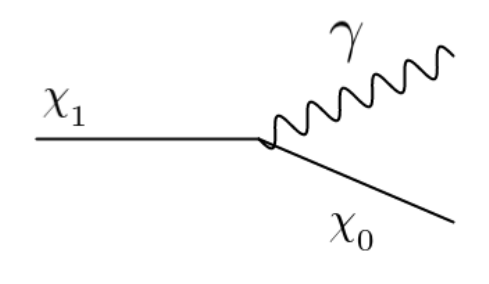}
\caption{ Feynman diagram describing two-body radiative decay of excited iDM state, $\chi_1$, into SM photon, $\gamma$, and lightest mode $\chi_0$. 
\label{FeynamDecay}}
\end{figure}

High-energy electrons scattering off nuclear targets offer a method for probing 
iDM, as depicted in Fig.~\ref{FeynamProduction}. Namely, this 
search  looks for missing energy in the reaction of electron scattering \cite{Gninenko:2013rka}, which arises 
when: (i) an off-shell Standard Model photon, $\gamma^*$, produces a pair of 
hidden fermions, (ii) a vector meson, $V=(\rho, \omega, \phi, J/\psi$),  
photoproduction followed by its decay~\cite{Barducci:2023hzo} into iDM particles:
\begin{equation}
e^- N \to e^- N \bar{\chi}_0 \chi_1.
\label{eNtoeNChi0Chi1}
\end{equation}

A crucial feature of iDM model is the mass hierarchy within 
the dark sector. Specifically, the masses of the two fermions are not degenerate but 
exhibit a positive splitting, quantified by the parameter 
\begin{equation}
\Delta = (m_{\chi_{1}} - m_{\chi_{0}})/m_{\chi_{0}} \gtrsim  0,
\label{SplittinDefinition}
\end{equation}
The presence of both the MDM operator (\ref{LagrangianMDM}) and the mass splitting~(\ref{SplittinDefinition}) 
renders the $\chi_1$ state unstable.  This allows $\chi_1$ to undergo a two-body decay, 
$\chi_1 \to \chi_0 \gamma$,  as illustrated  in Fig.~\ref{FeynamDecay}, 
producing, in principle, the single-photon semi-visible signature in NA64e. However, since for the  iDM parameter space considered in this work the energy of the 
decay photon is $E_\gamma \ll 50 $ GeV, which is a typical missing-energy threshold in the experiment, in the following we neglect the value of $E_\gamma$ 
and  assume that all the energy of the decay  $\bar{\chi}_0\chi_1$ pair from the reaction of 
Eq.~(\ref{eNtoeNChi0Chi1}) is carried away from the detector, thus producing the missing-energy signature in the experiment. The consideration of the possibility to combine both the single-photon and  missing-energy signatures which could enhance the sensitivity of the search is outside the scope of this work and will be considered elsewhere.  
\par Our further analysis  demonstrates  the NA64e's capability to explore the unconstrained area of the iDM parameter space
by using the missig-energy approach. Note, that  the NA64e ultimate number of the expected electrons on target (EOT)   is assumed to be 
  of the order of ${\rm EOT} \simeq 10^{13}$. Achieving the desired sensitivity, will require   
suppressing the background level in NA64e down  to  $\lesssim \mathcal{O}(10^{-13})$ per incident electron, which seems is quite possible \cite{NA64:2025nqq}.
Within this analysis, we examine exclusively the MDM interactions, given that 
the electric dipole moment 
operator~\cite{Masso:2009mu,Chu:2018qrm,Chu:2020ysb} yields a similar 
results. Furthermore, we don't consider the anapole moment and charge radius 
terms~\cite{Ho:2012bg,Gao:2013vfa,DelNobile:2014eta,Alves:2017uls}, as their 
parameter space is severely constrained~\cite{Jodlowski:2023ohn,Dienes:2023uve}.

This paper is organized as follows. In Sec.~\ref{Sec:Experimenta} we outline the  benchmark NA64e experiment.   In Sec.~\ref{Sec:SignalRate} we discuss the rate of 
iDM pair production at NA64e associated with bremsstrahlung like emission and 
vector meson decay. In Sec.~\ref{Sec:Sensitivity}, we derive the expected reaches 
of the NA64e fixed target facility in the  iDM  parameter space. 
In Sec.~\ref{ResultsSection} we provide a summary  and conclusion.

% \begin{figure*}[!t]
%\centering
%\includegraphics[width=0.48\textwidth]{feynman_DAP_brems.png}
%\includegraphics[width=0.25\textwidth]{feynman_GammaD_To_chi_chi.png}
%\caption{ 
%\label{MinimalALPportalBremsFeynman}}
%\end{figure*}

\section{The NA64e experiment
\label{Sec:Experimenta}} 

The NA64e experiment makes use of a 100 GeV electron beam delivered by the H4 beamline of the CERN  Super Proton Synchrotron (SPS). This beam has the maximum intensity close to $10^7$ electrons per spill of  $4.8,\text{s}$, and approximately $4000$ such spills are considered usable on a daily basis.

If the  iDM particles exist, they could be produced in the reaction of Eq.~(\ref{eNtoeNChi0Chi1}). 
 In this process, the electron beam with the energy $E_e =100$ GeV 
interacts with  heavy nuclei of an active target, which is  a lead-scintillator electromagnetic 
calorimeter (ECAL). Following the interaction, the $\chi_1$ state undergoes a rapid decay into $\bar{\chi}_0$ and a photon, 
$\chi_1 \to \chi_0 \gamma$. A considerable fraction of the initial beam energy $E_{\rm miss}$, typically $E_{\rm miss}\gtrsim 50$ GeV,  is carried away from the detector by the $\chi_1 \bar{\chi}_0$ 
pair, leaving no activity in a hermetic hadronic calorimeter located downstream of the ECAL target. The residual energy, 
defined as $E_{\rm e}^{\rm rec} = E_{\rm e} - E_{\rm miss}$, is measured as a deposit in the 
ECAL from the recoil electron in the reaction \eqref{eNtoeNChi0Chi1}. As a result, the presence of iDM  particles 
would be identified through an excess of events characterized only by a single
electromagnetic shower in the ECAL with the energy $E_{\rm e}^{\rm rec}$ , exceeding the expected  
background levels (see, e.~g.~, Ref.~\cite{NA64:2016oww}). 

There are several background sources in the NA64e experiment, which for the  number of accumulated ${\rm EOT} \simeq  10^{12}$ are rejected in total down to $b < 1$ events, 
for more details see, e.g., Ref.~\cite{NA64:2023wbi}. For the  exposure with ${\rm EOT} \simeq  10^{13}$, we assume that after the corresponding detector 
upgrade, the  expected number of  background events will  be $b < 1$ event \cite{NA64:2025nqq}. 

 \begin{figure*}[t!]
\centering
\includegraphics[width=0.7\textwidth]{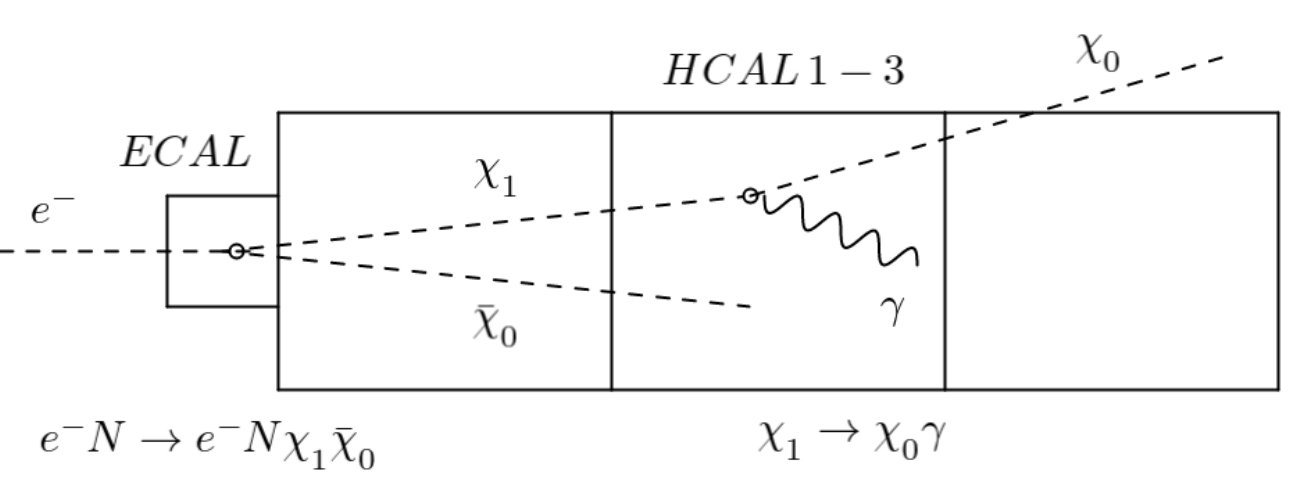}
\caption{ A schematic view of a missing energy signature \( V/\gamma^* \to \chi_1 \bar{\chi}_0 (\chi_1 \to \chi_0 \gamma) \)  produced after a 100 GeV \( e^- \) scatters off in the active dump, \( e^- N \to e^- N+ E_{\rm miss.} \). The \( \chi_1 \) particle  decays  within HCAL 1-3 into soft  undetectable 
photon $\gamma$ ($E_{\gamma} \lesssim 100~\mbox{MeV}$) and lightest sterile iDM state $\chi_0$. This 
sketch illustrates  the NA64e setup to search for  missing energy 
events~\cite{NA64:2016oww,NA64:2025nqq,NA64:2025ddk,Andreev:2021fzd,NA64:2021xzo,Banerjee:2019pds,NA64:2023wbi}. \label{NA64eDesign}}
\end{figure*}

\section{Signal rate
\label{Sec:SignalRate}} 

In this section, we discuss the production and missing energy signature of iDM  associated with the 
reaction~(\ref{eNtoeNChi0Chi1}), which is similar to the ones previously used by NA64e in their searches 
for dark photons at the CERN 
SPS~\cite{Gninenko:2017yus,Banerjee:2019pds,NA64:2021xzo,Andreev:2021fzd,NA64:2025ddk,NA64:2025nqq,NA64:2023wbi}.
Specifically, in our analysis we focus on the iDM production through two primary production mechanisms (see 
e.~g.~Fig.~\ref{FeynamProduction}): (i) the reaction $e^- N \to e^- N \gamma^* (\to\chi_1 \bar{\chi}_0)$ that 
yields iDM pairs via an off-shell photon, $\gamma^*$, analogous to bremsstrahlung 
emission~\cite{Arefyeva:2022eba,Gninenko:2025aek,Gninenko:2026mgn}; (ii) the exclusive photoproduction of a 
high-energy vector meson $V=(\rho, \omega, \phi, J/\psi)$ on a target nucleus $N$, via the reaction 
$\gamma^* N \to N V(\to\chi_1 \bar{\chi}_0)$. The initial photon $\gamma^*$ in the reaction originates from 
standard bremsstrahlung process, $e^- N \to e^- N \gamma^*$ (for detail see 
e.~g.~Ref.~\cite{Schuster:2021mlr,Gertsenberger:2025jed} and references therein).

As we discussed above, two-body decay, $\chi_1 \to \chi_0 \gamma$,
is kinematically allowed for positive mass splitting $\Delta \gtrsim 0$. 
Its partial decay width is given by~\cite{Jodlowski:2023ohn}:
\begin{equation}
 \Gamma_{\chi_1 \rightarrow \chi_0 \gamma} = \frac{\Delta^3 (\Delta + 2)^3 (m_{\chi_0})^3}{2\pi (\Delta + 1)^3 \Lambda_{\rm M}^2}.
 \label{DecWidth}
\end{equation}

\begin{figure*}[!tbh]
\centering
\includegraphics[width=0.495\textwidth]{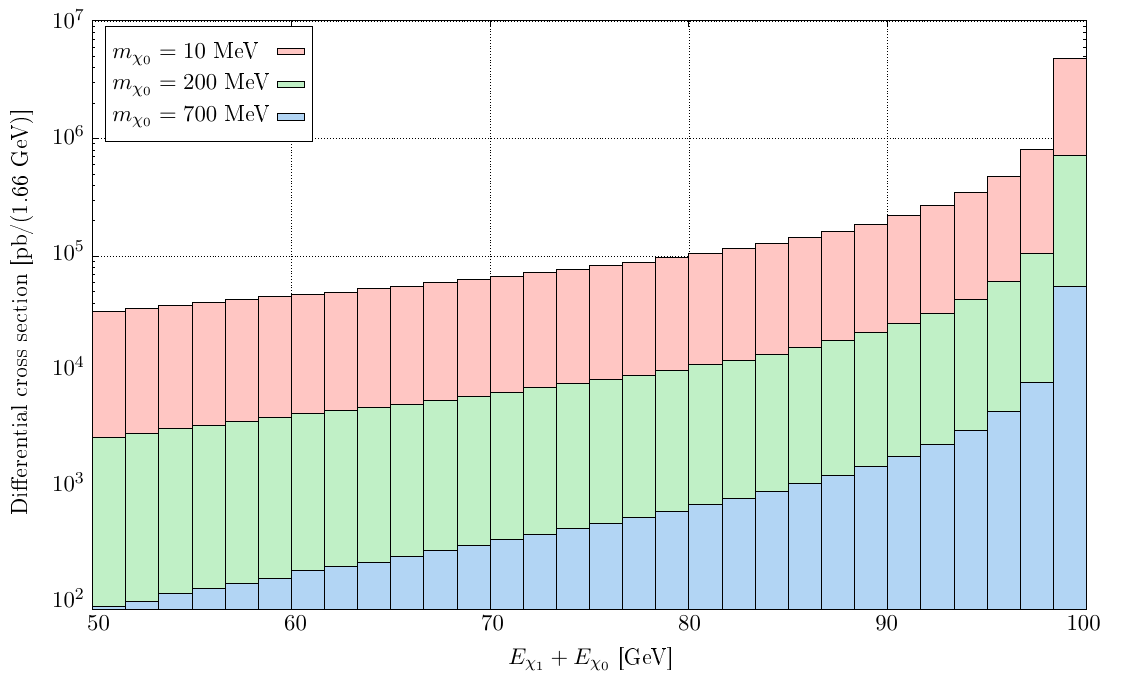}
\includegraphics[width=0.495\textwidth]{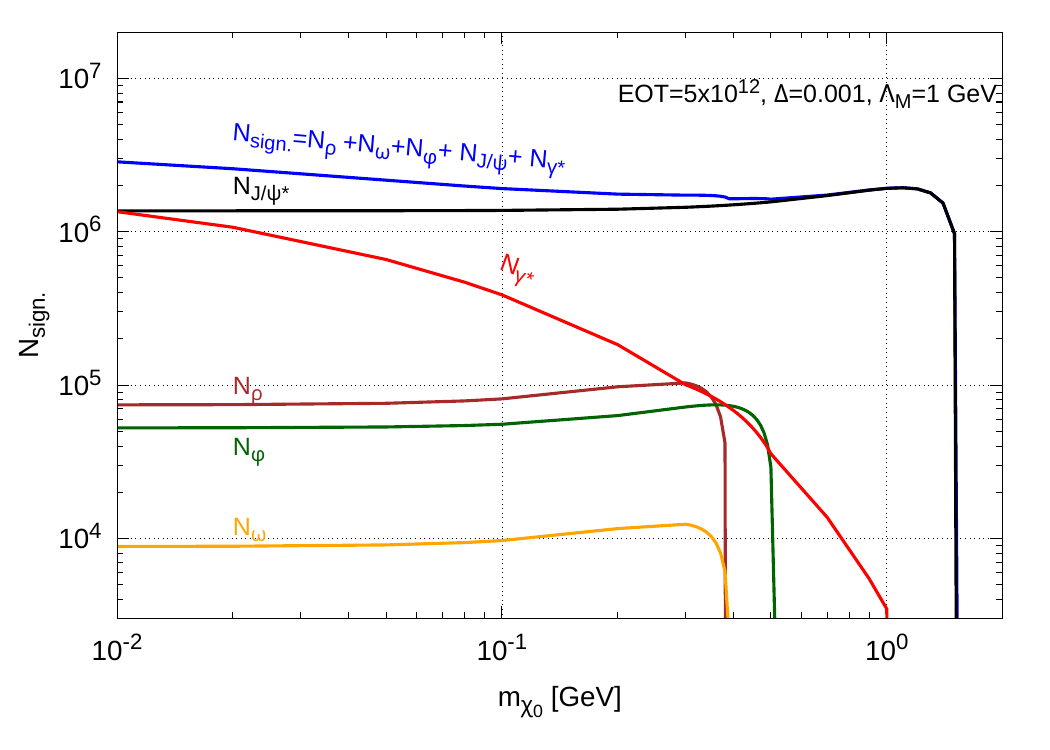}
\caption{ Left panel:  
the differential cross sections of the process $e^- N \to e^- 
N  \gamma^* (\to \chi_1 \bar{\chi}_0)$
as function of the  missing energy $E_{\rm miss}=E_{\chi_1}+E_{\bar{\chi}_0}$  
 for the set of iDM masses in the range $10~\mbox{MeV} \lesssim m_{\chi_0} \lesssim 700~\mbox{MeV}$.  
  The electron beam energy is $E_e \simeq 100~\mbox{GeV}$,  we  also set $\Lambda_{\rm M} =1~\mbox{GeV}$.
  Right panel:  the plot shows the  number of $\chi_1 \bar{\chi}_0$ pairs produced versus $\chi_0$ 
  mass for NA64e ($5\times 10^{12}$~EOT), with the splitting parameter  chosen to be 
  $\Delta=10^{-3}$. The hidden particles  are produced through the 
  bremsstrahlung-like emission $\gamma^*\to \chi_1 \bar{\chi}_0$ (red line), J-psi  meson $J/\psi\to \chi_1 \bar{\chi}_0$ (black line), rho meson  
$\rho \to \chi_1 \bar{\chi}_0$ (brown  line), omega meson $\omega \to \chi_1 \bar{\chi}_0$ (orange line), and 
phi meson decay $\phi \to \chi_1 \bar{\chi}_0$ (green line). The resulted number of 
$\chi_1 \bar{\chi}_0$ pairs produced at the NA64e is  shown by blue  line. 
\label{DiffCSAndNtotFig} }
\end{figure*}

The decay length of $\chi_1$ in the lab-frame  can be expressed for sufficiently small splitting $\Delta \ll 1$, as 
\begin{equation}
d_{\chi_1}\! \!\simeq\! 1.5 \text{m} \! \left(\! \frac{E_{\chi_1}}{10^2  \text{GeV}} \!\right)\! \!
\left(\! \frac{0.1 \text{GeV}}{m_{\chi_0}} \!\right)^{\!4}\!\!\! \left(\! \frac{0.05}{\Delta} \! \right)^3\!\!\! \left(\! \frac{\Lambda_{\rm M}}{10^3  \text{GeV}} \! \right)^2\!\!,
\label{DecLengthChi1}
\end{equation}
where $E_{\chi_1}$ is the $\chi_1$ energy. Note that 
Eq.~(\ref{DecLengthChi1}) implies that two body decay $\chi_1 \to \chi_0\gamma$ occurs within the  length of the NA64e calorimetric system of $\lesssim 6~\mbox{m}$.

As described in~\cite{Dienes:2023uve}, the energy of a photon originating from $\chi_1 \to \chi_0\gamma$ decay can be expressed in the lab frame as
\begin{equation}
E_\gamma = \frac{\Delta(2 + \Delta)}{2(1 + \Delta)}\! \left[ \gamma_{\chi_1} + \left[(\gamma_{\chi_1})^2\! - \! 1\right]^{1/2} \cos \theta_{\gamma\chi_1}\! \right]\! m_{\chi_0}, 
\end{equation}
where $\gamma_{\chi_1} \equiv E_{\chi_1}/m_{\chi_1}$ denotes the relativistic boost factor, and $\theta_{\gamma\chi_1}$ represents the angle, measured in the $\chi_1$ rest frame, between the photon's direction and the $\chi_1$ boost axis. 

Furthermore, to ensure that  the iDM  final states remain invisible within the 
 NA64e detector,  i.~e~. the energy of the decay photon is small, we require the mass splitting also to 
 be sufficiently small, $\Delta \ll 1$.  Following Ref.~\cite{Jodlowski:2023ohn}, we adopt two benchmark 
 splitting parameters:  $\Delta = 10^{-3}$ and $\Delta = 5 \times 10^{-2}$. 

Since the $\chi_1$ particle effectively inherits its boost factor from the decaying meson that produced it, and given that the characteristic energy scale for the meson is $E_V \simeq \mathcal{O}(10^2)~\text{GeV}$, the resulting boost factor is estimated to be $\gamma_{\chi_1} \propto E_V / m_V \simeq \mathcal{O}(10)$. Consequently, the characteristic photon energy from these decays within our parameter space is $E_\gamma \simeq \gamma_{\chi_1} m_{\chi_0} \Delta \lesssim \mathcal{O}(10^{-2})~\text{GeV}$. A single photon with such an energy cannot be detected by the NA64e calorimeter, as the energy deposition threshold, $E_{\rm th}$, is required to be sufficiently large, $E_{\gamma} \gtrsim E_{\rm th} \simeq  \mathcal{O}(1)~\text{GeV}$ (see e.~g.~Refs.~\cite{NA64:2021acr,NA64:2020qwq} and references therein for detail).

This parametric suppression of the potentially visible decay $\chi_1 \to \gamma \chi_0$ ensures that both dark matter fermions escape the experiment undetected, resulting in a clean missing-energy signature.

\subsection{Bremstrahlung-like production: $\gamma^* \to \chi_1 \bar{\chi}_0$}

The anticipated  number of $2 \to 4$ missing-energy signature  from the bremsstrahlung-like process 
(see Fig.~\ref{FeynamProduction}~(a)) can be estimated as:
\begin{equation}
N_{\gamma^* \to \chi_1 
\bar{\chi}_0} \simeq \mbox{EOT} \cdot \frac{\rho N_A}{A} L_T \int\limits_{E_{\rm min}}^{E_{\rm max}} dE_{\rm miss} \, \frac{d\sigma_{2\to4}(E_e)}{dE_{\rm miss}},
\label{NumberOfMissingEv1}
\end{equation}
where  \( \rho \), \( A \), and \( L_T \) denote the density, atomic mass, and effective length of the target, respectively. The symbol \( N_A \) represents Avogadro’s number. The differential cross section \( d\sigma_{2\to4}(E_e)/dE_{\rm miss} \) describes the \( 2\to4 \) bremsstrahlung-like process \( e^- N \to e^- N \gamma^* \to e N \chi_1 \bar{\chi}_0 \), and \( E_{\rm miss} = E_{\chi_0} + E_{\bar{\chi}_{0}} \) is the total energy carried away by the invisible $\chi_1 \bar{\chi}_0$ pairs. The integration limits \( E_{\rm min} \) and \( E_{\rm max} \) are set by the specific energy acceptance and selection criteria of the experimental setup. Specifically, candidate events are required to satisfy 
\(E_e^{\rm rec} \lesssim  0.5 E_{\rm e}\), 
corresponding to a lower integration limit of \(E_{\rm min} = 50\)~GeV in 
Eq.~(\ref{NumberOfMissingEv1}).

Additionally, we  assume that the electromagnetic shower in the ECAL is initiated  within its 
 radiation length of $0.56~\mbox{cm}$. This defines the effective interaction length 
in Eq.~(\ref{NumberOfMissingEv1}) as \(L_T \simeq X_0\), implying that the dominant 
production of hidden particles occurs  within the first $X_0$  of the  ECAL~\cite{Chu:2018qrm}. 
 
We compute the relevant $2\to4$ cross sections using the CalcHEP software 
package~\cite{Belyaev:2012qa}. We have extended the SM implementation in CalcHEP by 
introducing two new massive particles: the excited iDM state $ \chi_1$ and the 
lightest  $\chi_0$ mode.

The left side of Fig.~\ref{DiffCSAndNtotFig} shows how the differential cross sections 
change with the missing energy, $E_{\rm miss}$, within the signal box region, where 
$0.5 E_{\rm e} \lesssim E_{\rm miss} \lesssim E_{\rm e}$. These plots are shown for various dark matter masses $m_{\chi_0}$.
A clear peak appears in the NA64e data near $E_{\rm miss}\simeq E_{\rm e}$, meaning the 
cross section rises sharply in that range. This points to a strong forward focus when 
$E_{\rm miss} \gg m_{\chi_0}$, which implies that most of the beam energy goes into 
producing the inelastic dark matter  pairs. The numerical simulations reveal that the 
shape of the differential cross section depends weakly on $\Delta$, as long as we chose it 
sufficiently small, $\Delta \ll 1$. This also holds for the magnitude of the differential cross section.

\subsection{Production via vector meson decays: $V\to \chi_1 \bar{\chi}_0 $}

\begin{figure*}[t!]
\centering
\includegraphics[width=0.47\textwidth]{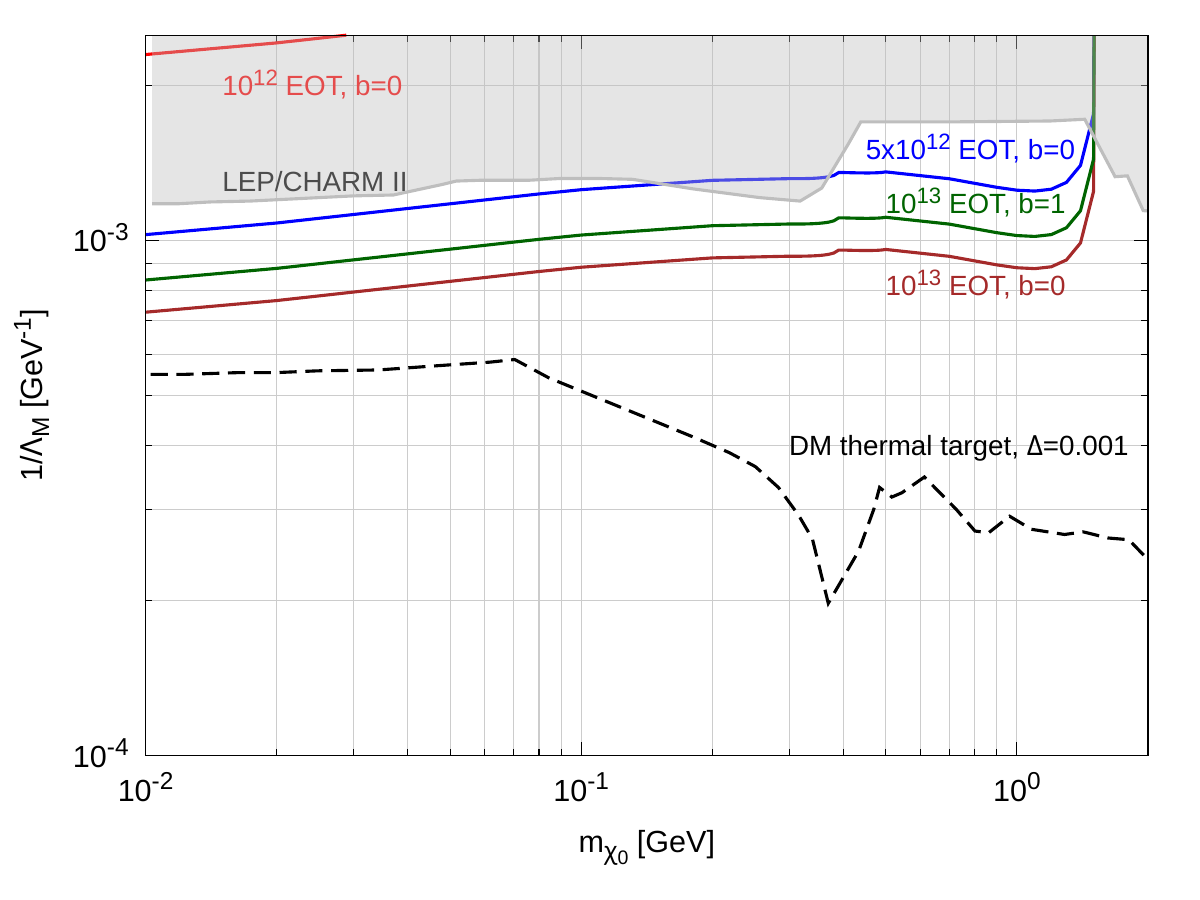}
\includegraphics[width=0.47\textwidth]{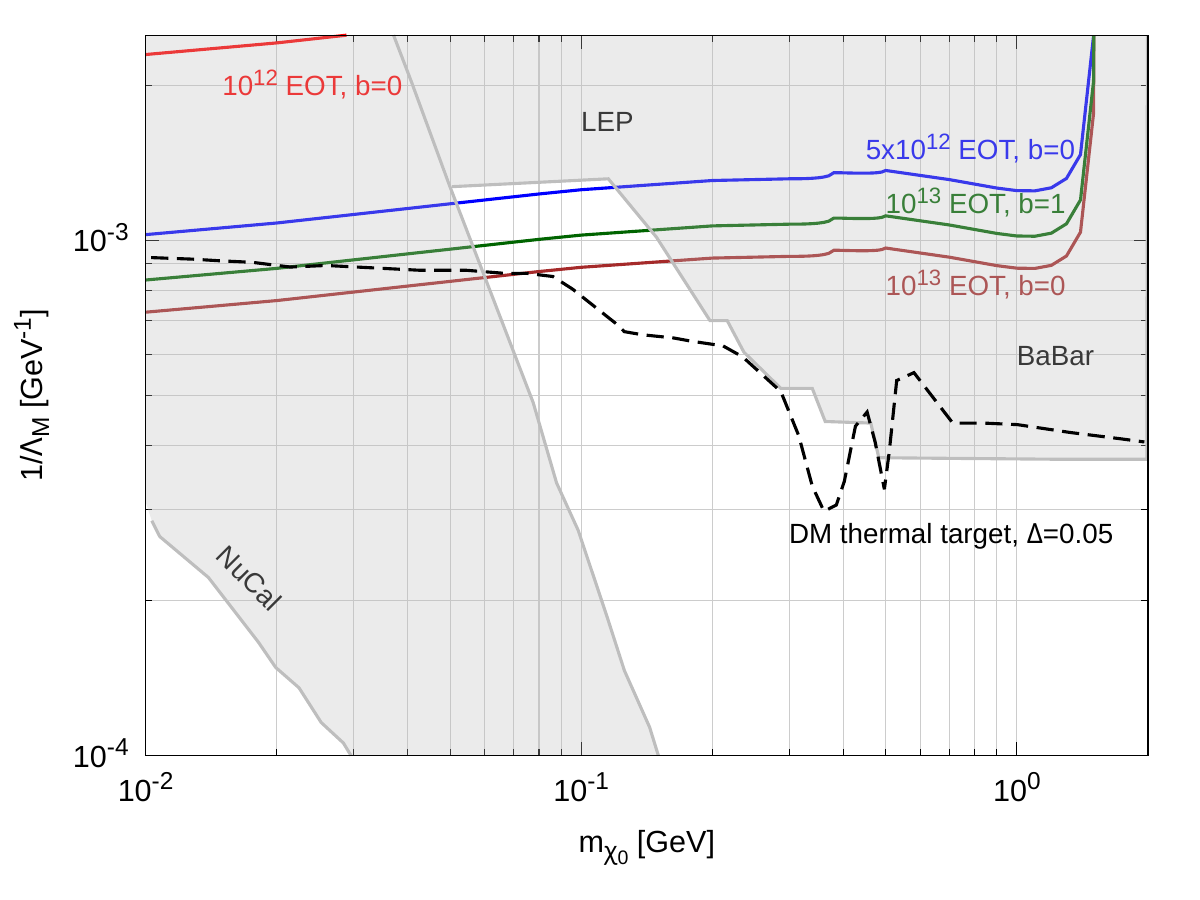}
\caption{Left panel: the expected limits at $90~\%~\mbox{CL}$ on $1/\Lambda_{\rm M}$ coupling for the NA64e fixed-target experiments as a function of  the lightest iDM mass $m_{\chi_0}$, implying mass splitting $\Delta=10^{-3}$.
Current limits of NA64e~\cite{NA64:2023wbi} are shown by red solid line and imply the typical statistics of $\mbox{EOT}\simeq 10^{12}$, these bounds have been ruled out by LEP~\cite{L3:2003yon,Fortin:2011hv} and CHARM~II~\cite{CHARM:1985anb,CHARM:1983ayi} experiments (shaded gray region). 
Blue solid line is the  projected sensitivity for NA64e experiment for $\mbox{EOT}\simeq 5\times 10^{12}$ and background free case, $b=0$. The expected reaches of the NA64e ($\mbox{EOT}\simeq 10^{13}$) for background free, $b=0$, and finite background case, $b=1$, are shown by brown and green solid lines, respectively. The iDM thermal target curve is shown by black dashed line, it  was adapted from Ref.~\cite{Jodlowski:2023ohn}.    Right panel: the same as for the left panel, but for  the benchmark mass splitting chosen to be as $ \Delta = 0.05$. 
The grey shaded region is ruled out from the BaBar~\cite{Izaguirre:2015zva}, NuCal~\cite{Blumlein:1990ay,Blumlein:2011mv}, and LEP~\cite{L3:2003yon,Fortin:2011hv} data, adapted from~\cite{Jodlowski:2023ohn}.  
\label{MinimalPortalFigExpectedReach} }
\end{figure*} 

The expected signal yield  from invisible vector meson decay into  an iDM   pair (see Fig.~\ref{FeynamProduction}~(b)) is expressed as 
\begin{equation}
N_{V\to \chi_1 \bar{\chi}_0} \simeq N_{V} \times \mbox{Br}_{V\to \chi_1 \bar{\chi}_0},
\label{NumbSignEVMesDec}
\end{equation}
where $N_{V}$ denotes the total count of the given vector meson~\cite{Schuster:2021mlr} for NA64e,
and $\mbox{Br}_{V\to \chi_1 \bar{\chi}_0}$ corresponds to its invisible branching fraction within the iDM framework~\cite{Jodlowski:2023ohn}.

The total amount of missing-energy signatures is then given by
\begin{equation}
N_{\rm sign.} \simeq N_{\gamma^*\to \chi_1 \bar{\chi}_0} + \sum_{V} N_{ V \to \chi_1 \bar{\chi}_0}.
\label{TotSignLDMX}
\end{equation}
The right panel of Fig.~\ref{DiffCSAndNtotFig} displays the total number of $\chi_1 \bar{\chi}_0$ pairs generated at the NA64e experiment as a function of~$m_{\chi_0}$.

For sufficiently light  masses, specifically $m_{\chi_0} \lesssim 100~\text{MeV}$, the NA64e signal from coherent $J/\psi$ photoproduction reaches a magnitude comparable to that of
$\gamma^*\to \chi_1 \bar{\chi}_0$.
The overall missing-energy signature count exhibits only a mild dependence on the  mass $m_{\chi_0}$, resulting in a relatively flat behavior across the sub-GeV region.

Although $\phi$ mesons are produced at lower rates than $\omega$ mesons, their invisible decays can nonetheless provide a larger signal contribution due to their enhanced branching fraction, as illustrated in Fig.~\ref{DiffCSAndNtotFig}.
Nevertheless, over the entire mass range of interest $10~\mbox{MeV} \lesssim m_{\chi_0} \lesssim 1.5~\mbox{GeV}$, the production of $\rho$, $\omega$, and $\phi$ vector mesons at NA64e remains subleading with respect to $J/\psi$ meson.

It is worth noticing from the right panel of Fig.~\ref{DiffCSAndNtotFig} that 
there is a kink in the NA64e  signal observed below the threshold  $m_{\chi_0} \lesssim m_{J/\psi}/2\simeq 1.5~\mbox{GeV}$  due to the contribution 
from the  $J/\psi$ vector meson decay to the total yield~\cite{Dienes:2023uve}.

The dominant contribution to the signal yield of the sufficiently heavy vector meson $J/\psi$ can 
be understood from the proportionality of the branching ratio to its mass 
squared, given that the number of such 
mesons produced is sufficient in the reaction $e^- N \to e^- N  V(\to \chi_1 \bar{\chi}_0)$ (see 
e.~g.~Ref.~\cite{Schuster:2021mlr} and references therein). Remarkably, an analogous statement holds 
for the heavy vector mesons  produced in proton beam dump 
facilities, $pp\to X  V (\to \chi_1 \bar{\chi}_0)$~\cite{Jodlowski:2023ohn,Dienes:2023uve,Chu:2020ysb}.

\section{expected sensitivities
\label{Sec:Sensitivity}}

In Fig.~\ref{MinimalPortalFigExpectedReach} we show by solid lines the expected reaches of NA64e solid lines. Specifically,
by using Eq.~(\ref{TotSignLDMX}) to calculate the expected number of \( \chi_1 \bar{\chi}_0 \) produced events, we derive projected exclusion limits on the portal coupling \( 1/\Lambda_{\rm M}\) within the minimal inelastic magnetic dipole DM framework. Assuming a 
background-free search, $b\simeq 0$, and a null observation, $n_{\rm obs.}\simeq 0$, we set a 90\% confidence level (CL) 
exclusion criterion corresponding to a signal yield of \(N_{\rm sign.} \gtrsim 2.3\) 
events for $\mbox{EOT}\simeq 5\times 10^{12}$. 

 To perform additional more conservative projection, in Fig.~\ref{MinimalPortalFigExpectedReach} we also 
 show   the expected reaches of NA64e implying the anticipated statistics of 
 $\mbox{EOT}\simeq 10^{13}$ for  both background free, $b=0$, and non-negligible number of regarding   events, $b\simeq 1$ (see, e.~g., Ref.~\cite{Gninenko:2025aek,Magill:2018jla} and references therein for detail).

In Fig.~\ref{MinimalPortalFigExpectedReach} we show by solid red line the excluded limits of NA64e for the current accumulated  statistics~\cite{Andreev:2021fzd} 
of $\mbox{EOT}\simeq  10^{12}$.  To be more specific, the current bounds of  
NA64e   reach the  couplings  at the level of  
$\Lambda_{\rm M} \gtrsim  2\times10^{-3} ~\mbox{GeV}$ for the  masses  of 
interest, $m_{\chi_0} \lesssim 1.5~\mbox{GeV}$, however these bounds have been already ruled out by other experiments for both benchmark splittings, $\Delta = 10^{-3}$ and  $\Delta = 5 \times 10^{-2}$. 
We  note that the projected limits of  NA64e for $\mbox{EOT}\simeq 5\times 10^{12}$ can rule out 
the typical couplings at the level of 
$1/\Lambda_{\rm M} \gtrsim  (1.0 - 1.5)\times 10^{-3}~\mbox{GeV}^{-1}$ for sub-GeV dark iDM masses.   

The thermal freeze-out production mechanism involving magnetic dipole iDM has been 
analyzed in detail in Ref.~\cite{Jodlowski:2023ohn}. In 
Fig.~\ref{MinimalPortalFigExpectedReach}, we show the typical thermal target curves for 
two specific benchmark splittings, $\Delta = 10^{-3}$ and $\Delta = 5 \times 10^{-2}$, 
adapted from Ref.~\cite{Jodlowski:2023ohn}. These lines imply that increasing the mass 
splitting value $\Delta$ shifts the relic curves upward relative to the constraints. As 
a result, a splitting of $\Delta = 10^{-3}$ spans sufficiently small couplings in the 
range $2 \times 10^{-4}~\mbox{GeV}^{-1} \lesssim 1/\Lambda_{\rm M} \lesssim 6 \times 10^{-4}~\mbox{GeV}^{-1}$, 
while the larger $\Delta = 5 \times 10^{-2}$ implies a more mitigated range for the coupling 
constant $10^{-3}~\mbox{GeV}^{-1} \lesssim 1/\Lambda_{\rm M} \lesssim 3 \times  10^{-3}~\mbox{GeV}^{-1}$. This means that NA64e is able to reach only a small fraction of the iDM thermal target 
curve, $1/\Lambda_{\rm M} \simeq 9 \times 10^{-4}~\mbox{GeV}^{-1}$, for $\Delta = 5 \times 10^{-2}$ 
and masses around $m_{\chi_0} \lesssim 100~\mbox{MeV}$, implying a sufficiently high number of 
electrons accumulated on target $\mbox{EOT} \lesssim 10^{13}$ and a background-free case.

\section{Summary and Conclusion
\label{ResultsSection}}

In the present paper, we have studied  the missing-energy signatures for the iDM production in the  electron  
fixed target experiment NA64e at the CERN SPS.  In particular, we calculated the production rate of the  magnetic-dipole 
 iDM pairs in the  reaction  $e^- N \to e^- N  \chi_{1} \bar{\chi}_0$. Based on this 
study, we have  determined  the expected sensitivity of NA64e to probe such particles with the 
anticipated statistics of $\lesssim 10^{13}$ electrons accumulated on target. 

 We show  that by accounting for heavy vector  mesons decays, $ \gamma^* N \to N V (\to \chi_1 \bar{\chi}_0)$ along with bremsstrahlung-like emission of iDM pairs, 
$e^- N \to e^- N \gamma^* (\to \chi_1 \bar{\chi}_0)$,  will allow NA64e to probe still unexplored 
region of iDM parameter space with sufficiently small splittings, $\Delta=5\times 10^{-2}$, and 
relatively light masses $ m_{\chi_0}\lesssim 100~\mbox{MeV}$. To achieve the required sensitivity 
an upgrade of the NA64e setup is currently in progress~\cite{NA64:2025ddk,NA64:2025nqq,Andreev:2023xmj}.

%\label{Sec9}

\begin{acknowledgments} 

We would like to thank  S.~Demidov, R.~Dusaev, A.~Pukhov, V.~Lyubovitskij and A.~Zhevlakov for very 
helpful discussions and   correspondences. The work of D.~K.~ on evaluation of the NA64e projected 
limits for the Dirac inelastic DM  was supported by RSF Grant No.~24-72-10110.

\end{acknowledgments}	

\bibliography{bibl}

\end{document}